\documentstyle [11pt] {article}
\makeatletter

\newcommand{\singlespacing}{\let\CS=\@currsize\renewcommand{\baselinestretch}{1}\tiny\CS}
\newcommand{\oneandahalfspacing}{\let\CS=\@currsize\renewcommand{\baselinestretch}{1.25}\tiny\CS}
\newcommand{\doublespacing}{\let\CS=\@currsize\renewcommand{\baselinestretch}{1.35}\tiny\CS}

\def\@citex[#1]#2{\if@filesw\immediate\write\@auxout{\string\citation{#2}}\fi
  \def\@citea{}\@cite{\@for\@citeb:=#2\do
    {\@citea\def\@citea{,\linebreak[0]\hskip0pt plus .2em}%
      \@ifundefined{b@\@citeb}%
      {{\bf ?}\@warning{Citation `\@citeb' on page \thepage\space undefined}}%
      \hbox{\csname b@\@citeb\endcsname}}}{#1}}

\newtheorem{rule-def}[theorem]{Rule}

\begin{document}

\newcommand{\la}{\lambda}
\newcommand{\si}{\sigma}
\newcommand{\ol}{1-\lambda}
\newcommand{\be}{\begin{equation}}
\newcommand{\ee}{\end{equation}}
\newcommand{\bea}{\begin{eqnarray}}
\newcommand{\eea}{\end{eqnarray}}
\newcommand{\nn}{\nonumber}
\newcommand{\lb}{\label}

\begin{center}
{\large {\bf A Village Astronomer:\\Life and Works of R.
G. Chandra}}
\end{center}

\begin{center}

Sudhindra Nath Biswas$^{1}$, Utpal
Mukhopadhyay$^{2}$ \& Saibal Ray$^{3}$

$^{1}${\it Mahatma Aswini Kumar Dutta Road, Nabapally, Barasat, North 24
Parganas, Kolkata 700126, West Bengal, India}

$^{2}${\it Satyabharati Vidyapith, Nabapally, Barasat, North 24
Parganas, Kolkata 700126, West Bengal, India\\ umsbv@yahoo.in}

$^{3}${ \it Department of Physics, Government College of
Engineering \& Ceramic Technology, Kolkata 700010, West Bengal,
India \\ saibal@iucaa.ernet.in}\\

\end{center}

\section{Introduction}
The ever flourishing human society always receives dedicated
service of a very few among its member for acquiring more accurate
knowledge about the phenomena of nature and thus with the
advancement of time the society enables to march forward in the
direction of prosperity. The name of late Radha Gobinda Chandra
(1878 - 1975) can be proudly enlisted along with many such
dedicated souls for his untiring service rendered in the cause of
development in astronomical knowledge with his humble might.

Chandra was born on 16 July, 1878 at the village Bagchar in
the house of his maternal uncle. The village Bagchar is situated
some three kilometer away from the district town Jessore.
Incidentally, Jessore, Nadia and Burdwan were contiguous districts
of Bengal belonging to as a state of undivided India. After the
independence of India in 1947, major part of the district Jessore
became a part of East Pakistan which again turned into a republic
country Bangladesh since 1971. His father Gorachand Chandra was
originally a resident of Burdwan district. After his marriage with
Padmamukhi$^1$, Gorachand remained as an inhabitant of his more
affluent in-laws.

In his maternal uncle's house, there prevailed a very congenial
environment for the all round nourishment of any growing child
like Chandra. His maternal uncle Abhaya Charan Dey was a
writer and as such he used to maintain a good library. It may be
mentioned here that in later years, Chandra had a free access
to a personal library, viz. `Bandhab Library' maintained by his
affluent friend Kedarnath Chandra who himself was a book
lover$^1$. Chandra's interest towards the cosmos was mainly
induced by Sarada Sundari Dhar, his maternal grand-mother. She was
a virtuous lady whose concept in astronomical phenomena was a
matter of envy to many. She could easily identify some of the
constellations, bright stars and the visible planets. Also, quite
a few learned family-friends such as Jogendranath Vidyabhusan
(1845 - 1904), the Editor of the Bengali magazine `Aryadarshan'
used to pay frequent visit to their house.

Like many other boys of middle-class family, Chandra started his
education at an early age of five years in a primary school of his
native village Bagchar. In course of time, he joined the Jessore
Zila High School to continue his secondary education. During his
studies in this school, there happened an incident that greatly
influenced the young mind of Chandra to quest for the knowledge
relating to cosmos. This happened in the process of
preparing his lessons as a student of class six. He had to go
through an essay entitled `Brahmanda Ki Prokando' (How Big the
Universe Is) written by a renowned Bengali writer Akshay Kumar
Dutta (1820 - 1886), in his Bengali text book `Charupath'. The dormant urge of
Chandra for acquiring knowledge on the wonder world of astronomy
was vigorously kindled by the text of the said essay. Since then
he intensified his studies on the object and also became more
serious for making observations of the celestial objects. However,
no reference of particular books which he did read or the
celestial objects which he observed during this period is still
unknown. At the age of 21 years, while he was still a school
student, in 1899, he got married with Gobindamohini Devi who was
the second daughter of Tribhangosundar Nath of Murshidabad
district$^1$. Unfortunately his academic career terminated with
the third futile attempt for qualifying at the Entrance
Examination conducted by the University of Calcutta for the
students of class ten.

Chandra spent two years after he had to give up his effort for
continuing formal education. A sense of self respect made him
impatient to get rid of his dependence on his family. So, after a
little quest, he got a service in the Government Treasury of
Jessore much to the dismay of his affluent family members. He was
employed as a `Poddar' (coin tester) in 1901 on a monthly salary of rupees
fifteen only. His duty as a Poddar of the Treasury was to examine
the genuineness of the metallic coins.

In absence of public transport system, Chandra had to travel daily, 
on a bicycle, all the way of three kilometer from his native
village to the Treasury. Throughout his entire service period, he
had been very sincere in discharging his duties at the Treasury
during day-time and simultaneously he also maintained his
responsibility of observing celestial objects in the role of a
world-class astronomer during night. Having completed the
continuous service of 35 years, he retired from the elevated post
of the Treasurer when he had been drawing a monthly salary of Rs.
175 and thus his day-time compulsion terminated.

After the partition of India in 1947, as also its state Bengal,
Chandra had to leave with his family and portable belongings, his
own house at Bagchar in erstwhile East Pakistan. He took shelter
at Sukhchar in the district of North 24 Parganas within the Indian
state West Bengal. Finally he settled in 1957 at Durgapally of
Barasat which at present is the Headquarter of the North 24
Parganas district. Here he spent a penurious life until his death
on 3 April, 1975.\\

\section{The Guru: Kalinath Mukherjee}
In a house, on the way to his working place Jessore Treasury,
Chandra used to pay regular visit after the days work was over.
There lived a renowned person named Kalinath Mukherjee, B. A., B.
L. (see APPENDIX I). He was practicing law in the District Court
of Jessore and was also known as an amateur astronomer of
considerable fame.

Chandra came to know about the interest and erudition of Mukherjee
in astronomy. He felt strong affinity for attending the discourses
on astronomy, arranged in the house of the latter. Somehow he
managed to have an access to the said discourses. Just after the
end of hard work for the day at the Treasury, he had to join the
discourses before returning home. He was tired by then and his
hunger for food was gradually becoming stronger. But his yearning
for the knowledge of astronomy was the strongest conviction to
absorb him into the discussion circle. Initially, an ordinary
person like Chandra used to receive cold reception from the host.
As he became a regular participant  of the discourses, his
importance was felt by Mukherjee himself and his associates. The
importance of Chandra to Mukherjee became so much that he was
entrusted with the responsibility of proof-reading for the books
of the latter. Even then he had to procure those works by
purchasing them from the market as he was not fortunate enough to
receive a complimentary copy from the author. Yet, Mukherjee was
his mentor and preceptor for inculcating astronomy in his young
mind.\\

\section{Observation Of Comets}
After taking part in the discourses on astronomical topics at the
house of Mukherjee, Chandra realized his limitations for the study
of the subject. Considering his own academic attainment, he
realized that it would be possible for him to achieve the
observational dexterity in astronomy rather than its theoretical
intricacy involving the concepts on higher mathematics and
physical sciences. Accordingly, he planned his future action and
began to observe the starry night sky with keen attention.
Gradually, he became acquainted with the constellations, Zodiac
and bright stars by the naked eyes. He later procured a binocular
to observe the still fainter celestial objects including the
meteors and comets.

Chandra observed a good number of comets by the naked eyes, as
also through the 3-inch and 6.25-inch telescopes acquired
subsequently in 1912 and 1928 respectively. He began his
performance as an amateur astronomer with remarkable observational
skill on the comet 1P/Halley 1909 R1. He made observations on this
comet by his naked eyes and also through binocular. This comet
remained visible from 25 August 1909 to 16 June 1911 while
passed its perihelion distance of $0.587208$~A. U.~\footnote{$1$~A. U. $=$
$149597870$ km.} on 20 April 1910 at $4^h~17^m~2.4^s$ U. T.

A report on his observation of the comet $1P/Halley$ was recorded
by Chandra in his book entitled `Dhumketu' (The Comet). At that
time Rai Bahadur Jadunath Majumder, Vedanta Bachaspati, M. A., B.
L., C. I. E. was the most revered person of Jessore. Rai Bahadur
inspired the people of the town so much that they became very
enthusiastic for the observation of predicted apparition of the
comet. By then, Chandra had no adequate experience for such
observation. Even he had no sky atlas except the `Bhagola
Charitam' prepared by Mukherjee. But, fortunately he received some
guidance from the two articles on the subject, published in the
`Probasi', a Bengali monthly magazine and authored by Jagadananda
Roy (1869 - 1933). Jagadananda was a renowned science teacher of
the school at Santiniketan under Viswabharati University, Bolpur
founded by the Nobel Laureate poet Rabindranath Tagore (1861 -
1941) who supplied a $4$-inch telescope to Jagadananda for
observation of the comet $1P/Halley$. By making personal contact,
Chandra wanted to know the predicted time and location of the
forthcoming comet from Roy who promptly fulfilled
his request. By the guidance of the latter Chandra became one
of the first observers to locate the comet $1P/Halley$ from India
on 24 April 1910 with the help of his binocular. He spotted the
comet for the first time as a small star-like object slightly below
the Venus and to the south of the star $\gamma$ Pegasi$^2$. The
comet's tail was scheduled to occult the Venus on May 2 1910, but
Chandra didn't observe the occultation although the Venus was
seen juxtaposed near the tail. He speculated that the transient
part of the tail was lying on the Venus and hence the occultation
was not visible. Afterwards, photograph of the comet $1P/Halley$
taken by John Evershed between $4.40^h$ and $5.10^h$~IST
revealed that the transient part of the cometary tail was indeed
lying on the Venus. This proves the high quality of intuition
possessed by the self-taught astronomer Chandra. However, he
observed the comet in its full bloom for the first time on May 10
at $3.20^h$~IST. According to him$^2$, the tail was passing
through the north of $\gamma$ Aquarii, west of $\alpha$ Aquarii,
north of $\beta$ Aquarii and south of $\epsilon$ Pegasi. He
observed some small stars through the tail to the north of
$\gamma$ and $\beta$ Piscium. The head of the comet was lying in
the second part of Pisces and the tail was extended to the last
portion of Capricornus. He continued his observations on the comet
which was not visible by naked eyes, maintaining exchange of
experiences with Roy of Santiniketan
($23^0~39^{\prime}$~N, $87^0$ $43^{\prime}$~E) and John Evershed (1864 - 1956), 
the Evershed effect discoverer, of Kodaikanal
Observatory ($10^0$ $14^{\prime}$~N, $77^0$ $28^{\prime}$ E)$^3$.

Chandra kept his vigilant eyes on the comet $1P/Halley,~1909~R1$
and subsequently published two articles in details on its
apparition as observed by him. Roy was very much impressed
by reading those articles and advised him to procure a telescope
in order to have better observations in future on the celestial
objects. The former already felt very much the need of such an
optical instrument and so his feeling was vigorously inspired by
the advice of the latter.

In the meantime in 1912, the then Government of India enhanced the
pay-scales of all of its employees. As a result, like all others
Chandra also received some amount of arrear money along with
his higher monthly pay. With this extra amount of money, he first
made an advance payment in April 1912 to the Bernard \& Co. of
England for purchasing a 3-inch refracting telescope. After two
months he took delivery of the telescope from the concerned
Shipping Transport Authority by paying the rest amount totaling a
sum of Rs. 160.63 (160 rupees 10 annas 6 pai). As the tube of the
telescope was made of card-board, so it became necessary to
replace it with one made of brass. For this assignment, the
Broadhurst \& Clerkson Co. of Calcutta were entrusted with at the
remuneration of additional Rs. 100. Thus, in view of his humble
monthly income, to acquire an apparently non-productive subject
like the telescope in exchange of a relatively large amount of
money must had been a courageous decision for a person like Chandra.

Chandra observed the comet
$7P/Pons-Winnecke$ 1927, which remained visible from 25 February
1927 to 10 January 1928 and passed the perihelion distance of
$1.039235$~A. U. on 21 June 1927 at about $1^h~34^m~10.56^s$ U. T. On 20
June 1927 at about $15.5^h$ U. T. ($21^h$ IST), he was busy with
his usual scheduled programme for observations of variable stars. 
He suddenly noticed a nebula-like object just North-West of
the bright star Vega. At that time the comet was visible on the
line joining star $\gamma$ Draconis and $\alpha$ Lyrae (Vega) and
was nearer (RA: $18^h~22^m~30^s$, Dec: $+ 40^0~30^{\prime}$) to
Vega. After consulting the handbook of British Astronomical
Association (BAA), he came to know that the object under his
observation was the comet $7P/Pons-Winnecke$. He observed the
comet until 7 July 1927. During the period of his observation, he
observed the comet to pass through the constellations Lyra,
Cygnus, Vulpecula, Delphinus, Pegasus, Acquarius, Sculptor and
Phoenix at a very fast speed of $40,000$ km/hr. In the context ``Search
for meteors from the Pons-Winnecke Radiant", the following report
was published in the journal `Nature'$^4$: {\it R. G. Chandra of Jessore, 
India, also reports a fruitless
search for meteors in the night of June 25. He states that Prof.
Ray of Bolpur saw two meteors radiating from the neighbourhood of
$\theta$ Bootes.} Here, Prof. Ray means Jagadananda Roy of 
Santiniketan as mentioned earlier.

The comet $2P/Encke$ 1927 having period of 3.30 years, the shortest 
among the periodic comets, was also observed by Chandra. He
searched out the said comet from the constellation Pegasus,
following the instruction of A. C. D. Cromlin, the Director of
BAA, and made observations until 17 January 1928. According to
him$^2$, he detected the comet Encke in 1928 at 7 PM from Jessore
with his 3-inch telescope in the Pegasus as a small nebulosity
near the Andromeda galaxy (M31). During the apparition, the comet
remained visible from 19 October 1927 to 3 April 1928. 

Chandra was not equipped with adequate data for determining the
location of a known comet. However, he succeeded to locate a long
period comet at a position $1^0$ to the south-west of the star
$\theta$ Ceti on 9 February 1941 with the help of his binocular.
The comet remained under his observation until 28 February 1941.
However, he could not identify the comet. Most probably it was the long
period comet $C/1941~B1~Friend-Reese-Honda$ with a period of 355
years. This comet remained visible from 18 January 1941 to 1
March 1941 as recorded in the Catalogue of Cometary Orbits-1999
(Catalogue 1999).

On 24 February 1943, at about $16.5$~U. T. ($22.00$ IST) Chandra
was engaged to observe variable stars with absorbed attention. All
of a sudden he noticed a nebula-like object near the star $\gamma$
Ursae Majoris. Later he could recognize the object as a comet.
According to Chandra's observation$^2$, the position of the
comet on that date (24 February 1943) was R. A. $11^h~55^m$, Dec
$+ 55^0$ at 10 PM. Its speed was slow and was visible in the east as a 
third magnitude star. The small tail was only visible through the telescope. 
He observed it as a bright nebula on the line joining the star
$\delta$ Ursae Majoris and $\gamma$ Ursae Majoris and was nearer
to the latter one. Though he could not identify the comet, yet it
is possible that he might have observed the long period comet
$C/1942~X1$ Whipple-Fedtke-Tevzadze because the `Catalogue-1999'
reveals that the lone comet remained visible during the period 17
November 1942 to 1 August 1943 was the comet $C/1942~X1$. He
made serious observations on the comet until 10 May 1943 with the
help of two refracting telescopes, one of his own 3-inch and the
other 6.25-inch lent him by the American Association of Variable
Star Observers (AAVSO). He recorded the apparent path and cometary
phenomena of the comet $C/1942 X1$ in details during the period of
his observations. He noticed the variations in its brightness by
measuring in magnitude scale on two occasions. First, on 8 March
1943, the magnitude of the comet reduced from the 5th to 4th and
again, on 16 March 1943 from the 5.5th to 5th. These observations
indicated that the cometary brightness instantly increased on
those two occasions. Later from the Journal of BAA, Chandra came
to know that the comet $C/1942 X1$ was really a variable one. Also
he realized from the said Journal that the phenomena of variations
in magnitude observed in the comet was due to influence of solar
magnetic disturbances during a sunspot maximum.

The systematic observations on comet were initiated in 1760 by
Charles Messier (1730 - 1817). Until 1999, astronomers all over
the world have discovered 1037 individual comets and observed to
make 1688 apparitions of these comets. An outstanding observer 
of comets amongst the contemporary observers from the Indian 
sub-continent Chandra observed a good number of comets. However, 
he could not discover any new comet like Elizabeth Roemer (born 1929) 
who recovered the highest ($79$) number of periodic comets! The only 
name of an astronomer from the Indian sub-continent associated with the
discovery of a comet and recorded in the `Catalogue-1999' is
Manali Kallat Vainu Bappu (1927 - 1982). Bappu jointly discovered
the comet $C/1949 N1$ with his teacher Bart Jan Bok (1906 - 1983)
and a fellow student G. Newkirk. This comet is known as the comet
$C/1949~N1~Bappu-Bok-Newkirk$.\\

\section{Discovery Of A New Star}
Chandra became interested 
for making observations on the planets, comets, meteors etc. 
His interest was so much intense that after
the acquisition of a 3-inch telescope, he gradually became an
expert observer of variable stars. In course of his routine-wise 
observation on stars from Bagchar ($23^0~10^{\prime}~5^{\prime\prime}$~N,
$89^0~10^{\prime}~15^{\prime\prime}$~E), Chandra had a chance in
1918 to locate a `New Star' which was actually a Nova. It was the
time for on setting of rainy season of the year. The light
radiating celestial objects were playing hide and seek behind the
running clouds of the sky. On the night of 7-8 June 1918, at
about $15.5^h$~U. T. ($21.00^h$~IST), he was watching the
celestial objects from a wide open place. Suddenly, he noticed
that the space around him was inundated with unusually bright but
smooth light! Also light with such intensity is visible only when
the bright Venus with $-4.4$ magnitude makes closest approach to
the Earth. He could not justify the reason for the appearance of
such unique brightness on a night before the New Moon when there
was no possibility for the appearance of Moon at the said hours of
night. Also, the bright planets such as the Venus and $-2.7$
magnitude Jupiter were not scheduled to rise above horizon at that
time in the night sky. Yet, he was at his wits end to explain the
presence of unusual brightness on the landscape at that night. As
the sky was infested with passing clouds, he did not intend to
make any observation in search of the source of light. Even
though, the suspicion that the source of light might be a
celestial object was gaining ground in his mind.

On the very next night of 8-9 June 1918, which was a New moon
night, at about $16.5^h$~U. T. ($22.00$~IST), Chandra easily
noticed a `bright star' at a glance. But the part of the sky under
his observation was covered with a veil of passing clouds. He
thought that the `bright star' might be the star Altair, the
brightest one of the Aquila constellation. After sometime, the
veil of clouds ran away and as a result both the `bright star' and
Altair became clearly visible with glare. The `bright star' (RA:
$18^h~44^m~43.48^s$, Dec: $+
0^0~29^{\prime}~28.2^{\prime\prime}$), in comparison, appeared to
be brighter than both the 0.77 magnitude Altair and $0.03$
magnitude Vega. These three stars were located in the same part of
the sky. Also, on a few successive nights, he observed the same
brilliance of light in the landscape and the `bright star' in the
sky.

A news about the `new star' in the constellation Acquila was
published in `The Statesman' on 12 June 1918. Chandra read the
news and realized that it was the same `new star' which he had
been observing since 7 June 1918. As it was at the time of early
stage of rainy season, the entire sky was covered with clouds
causing occasional rain ever since he made his observation on the
`new star' in the night of 8-9 June. It is due to this reason he
could not have any chance to repeat his observation on the star
until 16 June. In the night of 16-17 June 1918, during the
interval of $01.00^h$ to $01.50^h$ IST, the sky around the `new
star' remained clearly visible. The `new star' by then attained a
higher declination close to the Zenith, enabling Chandra to have a
good observation. Accordingly, he located the `new star' at the
south-west of $4th$ magnitude star $\theta$ Serpentis and
north-east of $3th$ magnitude star $\eta$ Serpentis. He also
estimated the brightness of the same star and found it to be
comparable to that of the $0.9$ magnitude Antares.

The sudden appearance of a `new star' to the naked eyes of Chandra
in the night of 7-8 June 1918 from a location where none other
observed before him was actually a `Nova' (in Latin the word `Nova' 
means `new'). This `new star' of the
constellation Aquila was designated by the astronomers as the Nova
Aquilae 1918. Incidentally, it is the first, if not the only one,
Nova whose spectrum at the pre-outburst stage had been recorded.
From the records it has revealed that it was an A-type blue star
of $10.5$ magnitude until 5 June 1918 but suddenly its brightness
flared up to $6$ magnitude after two days on 7 June. Just at the
very phase of changing the star of constellation Aquila into a
Nova, Chandra was the first, if not the lone, observer to notice
the transition phenomenon on the night of 6-7 June 1918 at
$21.00^h$ IST from Bagchar. His observation of the `new star' in
the following night of 7-8 June at $15.5^h$ UT, even at a higher
brightness of 0.03 magnitude equaling that of Vega, confirmed the
Nova formation of the star. Although the star flared up to a
maximum of $-1.4$ magnitude brightness it eluded him due to
unfavorable sky condition.

A report on the first observation of the `new star' was authored
by Chandra in a widely circulated Bengali monthly magazine the
`Probashi' in its Sraban (July-August, 1918) issue of the same
year. Jagadananda Roy from Santiniketan read this article on the
observation of the `new star'. Understanding the importance of his
observation, Roy advised him to send the report to Edward
C. Pickering (1846 - 1919) of the Harvard College Observatory who
might appreciate his work and properly ventilate before the
astronomical community. Although Chandra was otherwise busy, yet
he did not realize the importance of his work, nor did he know the
formalities of reporting the discovery of a Nova. Incidentally, he
was very late to report the discovery and subsequently missed the
credit of becoming the discoverer of the Nova Aquilae 1918 (see
APPENDIX II). However, Campbell realized his agony for missing the
credit of being the discoverer of the Nova 1918. In a letter dated
24 June 1921, Leon Campbell (1881- 1951), the Chairman of
Telescope Committee, AAVSO, encouraged him by writing - {\it ``You
have taken up the Nova search work in a good spirit, and I hope
you may be rewarded some day with a real Nova discovery"}$^1$.\\

\section{Variable Star Observer}
Though the report of Chandra on the
observation of the Nova Aquilae 1918 reached six months late to
Pickering, the pioneer astronomer of star classification, yet the
later was impressed by the work of the former. In order to
appreciate Chandra's observational work, Pickering sent him a few
valuable books, star map, Revised Harvard Photometry of Stars and
some published works on the Nova Aquilae. Subsequently, Chandra
was elected as the honorary member of learned societies like the American
Association of Variable Star Observers (AAVSO), British
Astronomical Association (BAA) and Association Francaise de
Observteur (AFO), Lyon, France. Instead of paying any subscription, as
a member of these associations, his responsibility was to
contribute his collected observational data on the periods and
instant magnitudes of the variable stars. With the help of these
data, the professional astronomers were able to identify the
physical characteristics of the stars, in particular, variable
stars. Since 1919, his observational data were published in the
Monthly Report of AAVSO$^5$, Memoirs of the BAA$^6$ etc.

The site from where Chandra used to collect data by making
observations on stars was located at Bagchar, a remote village in
the Eastern India while the majority of the other variable star
observers were stationed either in Europe or in America. So, to
the astronomers of these Western countries, the data sent by
the lone observer from the Eastern longitude, were of very much
importance. In spite of his different type of professional
engagement in a Government Treasury, he not only showed keen
interest to, but also devoted tremendous labour and responsibility
for the regular observation of variable stars. The axes of both 
the eyepiece and objective lenses of his own
3-inch telescopetelescope were collinear. So to observe the objects
around the zenith region, with the help of such a telescope, was
physically very strenuous work~\footnote{Incidentally, the first author
experienced, how difficult is in making observations on stars with
the help of this particular telescope while he was requested to
keep it under his care from 1989 to 1996.}. Overcoming all these
constrains, Chandra was able to measure the brightness of several
hundreds of stars in a single month with the said telescope. His
observational data, as it was admitted by the contemporary
professionals, were not only quantitatively rich, but also
qualitatively excellent. For this achievement, time and again he
received letters of appreciation from highly esteemed persons of
the stature of Harlow Shapley (Director, Harvard College
Observatory), Leon Campbell (Recording Secretary of AAVSO),  Felix
de Roy (Director, Variable Star Section, British Astronomical
Association) etc. (see APPENDIX III).

The members of AAVSO were so much impressed with his work as an
observer of variable stars, in particular, that they
decided to lend Chandra a more powerful telescope than the one he
had. Accordingly, Leon Campbell (1881-1951), the then Chairman of
Telescope Committee of AAVSO, offered a proposal for lending a
6.25-inch reflector telescope through the letter dated 12 August
1924 (see APPENDIX IV).  It may be mentioned here that when the
said telescope reached Calcutta, Nagendranath Dhar, owner of the
telescope manufacturing company `Dhar Brothers' took the trouble
of bringing the telescope from the railway station. Not only that, 
Dhar carefully made an equatorial 
stand for the telescope and presented it to Chandra$^1$.

It is difficult to collect and represent the huge number of
observations made by Chandra throughout his life. One can realize 
from the available record about the magnitude of extra labour he
had to exert as an observer. In the year ending in
October 1926 he made no fewer than 1685 observations, of which he
made 226 observations in the month of March alone$^7$. Also
available the number of observations he made on the 34 individual
variable stars during the period 1920-24 and reported to the BAA$^8$
(Table-1).

\begin{table}
\caption{Observations made by Chandra on 34 variable stars (1920 -
1924)} 
\centering
\bigskip
{\small
\begin{tabular}{@{}llrrrrlrlr@{}}
\hline \\[-9pt] VARIABLE OBSERVED & NO. & VARIABLE OBSERVED &
NO.\\ \hline \\ [-6pt]

R Andromaedae &  21& V Cygni   &14\\

W Andromedae &55& R Draconis &16\\

R Aquilae &7 & T Draconis &3\\

R Arietis &42& R Geminorum &33\\

R Aurigae   &25& S Herculis &3\\

X Aurigae  &50&T Herculis &34\\

R Bootis  &65&U Herculis &45\\

S Bootis &53&R Hydrae &101\\

V Bootis &57&R Leonis &33\\

R Camelopardalis &21&U Orionis &48\\

X Camelopardalis &29&R Pegasi &7\\

T Cassiopeiae &39&R Serpentis &10\\

T Cephei &17&V Tauri &18\\

s (Mira) Ceti &136&R Ursae Majoris &30\\

S Coronae &62&S Ursae Majoris &69\\

X Cygni &86&T Ursae Majoris &48\\

R Cygni &19&S Virginis &52
\\

\hline
\end{tabular}
}
\end{table}

During the period of his long observational activities, he was
able to report as many as 37215 results on variable star
observation to various astronomical associations from 1919 to
1954$^7$. As a mark of respect to the valuable observations by 
Chandra, in a letter dated 3 March 1928, Leon Campbell expressed
his gratitude with the following words - {\it ``The results which you are
obtaining on the variables are excellent and we class you as one
of our best contributors"}.\\

\section{Interactions with Professionals}
In the process of variable star observations Chandra came in
contact with many professional astronomers of America, Europe and
India as well. It appears from the correspondence made to him,
from time to time, that those professionals used to treat him as
one of their respected colleagues. Among those professionals, Leon
Campbell who was the Recorder and subsequently became the
President of AAVSO, was very close to him. From his letter 
(dated 3 March 1928) it reveals that in spite of taking best care,
the 6.25-inch telescope lent to Chandra by AAVSO, reached in a
damaged condition. So, embarrassed Campbell who was the main
architect in lending the same, wrote him how the damaged parts of
the telescope could be repaired in India, or as an alternative
measure, requested to send back at Harvard Observatory for the
same work. Considering the enthusiasm and eagerness of Chandra to
use a more powerful telescope than his own 3-inch, Campbell 
through the same letter, inspired him by the following words:

{\it ``The delay incidental to such an unfortunate thing is to be
regretted, and I hope that you can soon have the six inch working
as you have long hoped"}.

Chandra also observed the occultation and lunar eclipse of 20
February 1924 and its report was published in The Journal of the
British Astronomical Association$^9$. That report goes like this:
{\it Arrangements were made with two friends to observe the Lunar
eclipse and occultations of stars, one to watch the minute hand,
the other to watch the second hand, and both counting the minutes
and seconds independently and record the time when I shouted
`one', `two' and 'three' from the telescope. This was carefully
done so that we get a very accurate time. Time was taken from the
Jessore Telegraph Office at 4 p.m. at which hour each day the time
is signaled from the Government Telegraph Office at Calcutta. The
sky was very fine and seeing very good: the observations were made
with naked eye, with binoculars and with a 3-inch refractor using
powers of 32 and 80.}

About the observation of lunar eclipse of 8 December 1929 Chandra 
reported it to Willarad J. Fisher of Harvard College
Observatory with some quarries. Fisher replied to his all quarries
in a letter dated 30 January 1930, starting with these lines:

{\it ``I have yours of December 14 with a very nice description of
lunar eclipse of December 8, as observed by you at Calcutta or
neighbourhood. I am greatly obliged for this. I also note your
quarries about phenomena observed. To these I can return only in
complete answers ..."}

The report of the observation of Lunar eclipse of September 26
1931 was published in {\it The Journal of the British Astronomical
Association}$^{10}$ and observation of Annular Solar eclipse of
21 August 1933 was also reported in the same journal$^{11}$. His
short communication on `Rahu' was published in the above mentioned
journal$^{12}$ in a different issue.

From a letter dated 28 August 1928, written by A. N. Brown of BAA
we come to know the regularity with which Chandra used to
make observation on variable stars and reported on results to the
Association. A part of the said letter is quoted here:

{\it ``I acknowledge with thanks the report of 15 complete sheets
of your observation of variables made this year ... I have so far
only just glanced over your sheets, but this glance is sufficient
to show that you have again done valuable work, particularly,
perhaps, in the regularity of your observations of some of the
Irregular $U$ Germinorum etc. in spite of the unfavourable weather
with which you say you have had ..."}

The reaction of Y. M. Holborn, the then secretary of BAA, implies
how indispensable were, for the professionals, the data collected
by Chandra from the observation of variable stars. When the latter
tendered his resignation from the membership of BAA, then the
former reacted as follows (letter dated 30 January 1941):

{\it ``I am passing on your letter of resignation to Mr. Brown who
deals with these things. But I must say, I think it is a great
pity to resign at this time when the Association is in the utmost
need of support.

Your longstanding work for the variable star section too will be
greatly missed just at the time when Lindley and others like
myself with full time war duties have had to give up observing.

I beg to you as an old member to reconsider this decision of
yours."}

The resignation must have been tendered by Chandra due to his old
age. Still it is not known whether the request of Holborn was
complied or not.

But, the request of M. K. Bappu was complied by Chandra. The
request by Bappu to get 3-inch telescope of Chandra as
a loan for observation for a brief period. Eventually, Bappu
expressed his gratitude in a letter (14 August 1945) by the following words:

{\it ``It is very kind of you to offer me the loan of your 3-inch
refractor with its accessories so as to enable me to continue my
observations of variable stars and I thank you heartily for the
same. I am also grateful to Prof. Campbell for kindly recommending
me to you."}

For this generous act, Campbell also thanked Chandra through a
letter (2 February 1946):

{\it ``It is certainly generous of you to place on loan to Mr.
Bappu the three inch telescope and I thank you on behalf of the
Association as well as on my behalf.

Mr. Bappu was an excellent observer when he had to access a large
telescope and I am looking forward to future. You might be
interested to know that his son is also very much interested in
variable star observing, and to date has been contributing
observations made with the naked eye."}

Due to his advanced age Chandra became incapacitated to use 
the 6.25-inch telescope and intended to return it to AAVSO, 
in compliance with the condition of its Telescope
Committee. But the authorities instructed him to transfer the
telescope to M. K. Vainu Bappu, the son of M. K. Bappu., instead
of returning it back to AAVSO. Vainu Bappu received the telescope
in 1958, while he had been working as the chief astronomer, Uttar
Pradesh State Observatory (UPSO), Nainital~\footnote{This is now known as
the Aryabhatta Research Institute of Observational Sciences (ARIS).}. After
working at the UPSO for the period 1954-1960, Vainu Bappu came to
Kodaikanal observatory in 1960 as its Director. Latter on, in 1971 he founded the
Indian Institute of Astrophysics (IIA) at Bangalore. The
said 6.25-inch telescope has been installed at the very entrance
of Cavalur Observatory ($78^0~50^{\prime}~E, 12^0~35^{\prime}~N$)
which is under the direct supervision of IIA. A letter of Prof. J.
C. Bhattacharya reveals that M. K. V. Bappu had a tremendous
respect for Chandra and time and again showed this telescope to
young astronomers who came to work at IIA as a mark of
perseverance and self determination of its user, viz.
R. G. Chandra$^7$.\\

\section{The Legendary 3-inch Telescope}
As his age advanced, Chandra began to wind up his involvement in
astronomical activities. In 1959 he donated his all books and
journals on astronomy together with his legendary 3-inch telescope
to a neighbouring Higher Secondary School, Satyabharati Vidyapith,
Barasat, Kolkata. During his lifetime, the telescope was rarely
used for observation by the School authority except on a few
occasions. Although some of the authorities, teachers and students
used to offer felicitation to Chandra on the occasion
of his every birth anniversary.

Around the period 1970, in connection to the Naxalite movement,  
a section of misguided youths started to
destroy the office and library of the schools in West Bengal. In
one such raid, many astronomical journals and valuable books of
the said School, including those donated by Chandra, were destroyed.
However, apprehending an untoward incident to occur, the
authorities of the School had the wisdom to shift the 3-inch
telescope in a outside safe custody. Thus, the legendary telescope 
was saved from the devastation. Chandra had to bear such terrible 
shock before he died on 3 April 1975.

The said legendary telescope remained out of the premises of the
School for a long time. The first author, having completed sometime
earlier a senior fellowship under the University Grants Commission
scheme for writing University level text books, joined the
School on 1 September 1981, as an Assistant Teacher of
mathematics. Around the time the comet $1/P~Halley$ made its
perihelion passage on 9 February 1986, he was requested by the
International Halley Watch organization to report on its
observation. Fortunately, the Department of Applied Mathematics,
Calcutta University, offered the service of their 3-inch refractor
to the first author, enabling him to guide the students and teachers of
both the Calcutta and Visva-Bharati Universities for their
observation of the comet. But he was unable to use the legendary
telescope of the School for the observation of the comet
$1/P~Halley$, as it was still in the possession of safe custody.

However, due to concerted effort of the authorities and teaching
staff, along with the generous cooperation of Prof. Amulya Bhusan
Gupta, it was possible to bring back the said telescope under the
supervision of School by the end of 1986. At that time, the
comet $1/P~Halley$ was receding to distances much beyond the reach
of the 3-inch telescope.

During this time, Dr. Ranatosh Chakraborty, a college teacher and
inhabitant of Barasat, took initiative in ventilating the
achievements of Chandra as an observational astronomer, by writing
book and articles in both the national and foreign journals and
also propagating talks through the radio and television
transmissions. He was instrumental in founding the `Radhagobinda
Memorial Society', Barasat, which organized his several birth
anniversaries starting from 1986 at different local institutions
as also in Kolkata. On almost every occasion the legendary
telescope was displayed for public viewing. Later, the telescope
was also displayed at the Birla Industrial and Technological
Museum, Kolkata in 1989 and again on the occasion of the Fourth
All India Amateur Astronomers' Meet at Presidency College, Kolkata
in 15-16 January 1994.

In the meantime Kalidas Chandra, the eldest son of Chandra,
requested School authorities for the proper maintenance and use
of the telescope donated by his father. As an effect the first 
author was entrusted with those responsibilities
from 13 April 1989. Having received the legendary telescope with
deep reverence to the historical instrument, he carefully made all
the component parts cleared and suitable for making observations.
Since then, the telescope was displayed at School and other
institutions for making observations on every possible celestial
event like the eclipses, planetary transits over the solar disc,
meteor showers and so on. After his retirement the author returned
the telescope once used by Chandra with all its components intact
to the then Teacher-in-Charge of the School on 27 June 1996.\\

\section{Calendar Reformer Chandra}
An ardent astronomer Chandra noticed the age old enigma in the
Indian calendars known as Panchang or Panjika. The enigma was that
the timings predicted in the contemporary Panjikas for the
occurrence of celestial events were not in conformity to the
actual observations. In some cases, such enigma prevails even
today. As the people are supposed to observe their respective
religious rituals depending on the predictions from Panjikas, so
the anomaly in timings always creates confusion among the
populace. Most of the Panjika makers were accustomed to use the
reference elements set in the Siddhanta Jyotisha calendar 
reformed way back in 400 A. D. At that time, ancient Indian
astronomers formulated some astronomical principles to determine
and successfully predict in their almanac the correct timings of
the celestial events. These events are the markers for observance
of religious rituals besides the day to day activity of the
people. They also realized that the reference point for the
calculation of time, termed as the Vernal Equinox, is not fixed.
So they advised the future Panjika makers to make adequate
adjustment in the timings from time to time in order to maintain
correctness of their predictions.

Since the concept of gravitation and its celestial manifestations
were yet to know at that time, so the ancient Indians could not
explain how the Vernal Equinox recedes. At the advent of
gravitation it was realized, in 1687, that the precessional motion
of the Earth is mainly responsible for the lagging of equinoxes
called the precession of equinoxes. Subsequently, it was also
found that the Vernal Equinox recedes towards the west along the
ecliptic to the extent of $50.28^{\prime\prime}$ causing a delay
of $20^m~24.32^s$~per year, for the occurrences of all celestial
events on the Earth.

The successive Panjika-makers of India maintained their
traditional orthodox attitude in preparation of their almanacs on
the basis of timings set during the Surya Siddhanta period. Surya
Siddhanta is an astronomical compilation, the writing of which
started from the Fourth century. They thought that as because
these timings were set by ancient sages, so it must be pure for
the purpose of religious observances and thus they deliberately
paid no heed to the advice for adjustments. Such orthodox practice
prevailed in the Seventeenth Century, even after the realization
of the precession of equinoxes and also continued to the Twentieth
Century! As a consequence, the predicted timings of the celestial
events published in the Panjikas, gradually continued to recede
from the observed ones by larger interval of time than ever.

There were public demands, from time to time, to make the timings
predicted in the Panjikas be corrected, as very often those were
not correlated with the actual observed ones. During the later
part of the Nineteenth Century, the erudite people from all over
India, like Lokmanya Bal Gangadhar Tilak (1856-1920), Pandit Madan
Mohan Malabya (1861-1964) of Kashi, Mahamohapadhyay Chandra Sekhar
Singh Samanta (1835-1904) of Orissa and Acharya Jogesh Chandra Roy
Vidyanidhi (1859-1956) of Bengal began to advocate for the reformation 
of Indian calendars.

In contemporary Bengal there were two Panjikas of which one was
Gupta Press Panjika published from Kolkata and the other Kalachand
Panjika from Sreerampore. Among many others Sri M. M.
Bandyopadhyay, a Jamider of Talinipara, also pointed out the
discrepancies contained in these Panjikas and suggested for the
preparation of corrected ones. In compliance to the suggestion,
Sri Madhab Chandra Chattopadhyay (1829-1905), a retired engineer
began to publish Visuddha Siddhanta Panjika (VSP) from 
1890~\footnote{Corresponding Bengali era 1297} 
on the basis of British Nautical
Almanac (BNA). The BNA was prepared on the basis of actually
observed motions of the celestial objects and considering their
effect due to the precession and nutation of the Earth on their
timings. Thus, the predicted timings of the celestial events
published in VSP had been in close conformity with their observed
ones.

It was not possible to elucidate the vast populace about the
astronomical aspect of Panjika making, nor were they able to come
out from the clutches of age-old traditional belief. So the newly
reformed almanac could not gain enough popularity and as such
financial patronage to compete with the traditional Panjikas. At
this juncture of time Chandra began to write articles, in the well
circulated journals, on the discrepancies observed in the
predicted timings of the traditional Panjikas. As a result, the
traditional Panjika-makers cleverly began to use the timings from
BNA, for predicting the occurrences of such celestial phenomena as
the eclipses, New and Full Moons, since those celestial events
could be easily verified by naked eyes. But for predicting the
other celestial events like the Tithis and Nakshatras they
maintained their devotion to the traditional method of
calculation. In this manner, in the name of religious purity they
continued predicting wrong timings in their publications to the
extent of 5 to 6 hours as because these timings cannot be verified
by the common people without having advanced knowledge in
astronomy and observational dexterity.

In his relentless effort for the reformation of Indian calendars,
Chandra not only published many articles like the one in
October-November, 1927 (Kartik, 1334) issue of Prabasi, a Begali
monthly magazine, but also joined the `Jyotish Parishad' as a
member. The Jyotish Parishad was founded by Indra Nath Nandi as a
learned organization in 1930-31 (1337 Bengali era) with an
objective of popularizing astronomy in general.

But, in particular, the members of the organization became very
active to achieve success in the effort of the calendar
reformation as their contemporary subject. On 22 December 1936,
the members of the Jyotish Parishad observed the `Chandra Sekhar
Day' in commemoration of the birth centenary of Chandra Sekhar
Singh Samanta, whose date of birth was 13 December 1835. Chandra
Sekhar was an eminent astronomer and an active reformer of
traditional calendars. In the said centenary celebration which was
held in the premises of Sanskrit College, Kolkata, Chandra
presented an erudite article on the reformation of contemporary
Panjikas. His said article was so impressive that subsequently it
had been published in the Education Gazette for its wide
circulation.

With the passage of time, the movement for reformation of Indian
calendars had been gaining momentum. As a result, immediately
after the independence, Government of India appointed in 1952 the
Calendar Reform Committee (CRC) under the Chairmanship of world
famous astrophysicist Prof. Meghnad Saha, F. R. S. (1893-1956). On
invitation by the CRC, Chandra placed his suggestions through a
letter dated 3 April 1953. The summery of his suggestions as
printed in the report of CRC are given below:\\
                  (i) Advocate `Nirayana' system of
                  calculation;\\
                 (ii) Initial point to be taken $180^0$ from the star
                 Spica;\\
                 (iii) Correct calculation to be adopted in the
                 calendar;\\
                 (iv) 21 March should be called as `Mahavisuva Din' and not `Mahavisuva
                        Samkranti'.

The CRC committee submitted its report with recommendations in
1955 to the Government of India. As per recommendations, the
Government started work for preparation of the Indian Ephemeries
and Nautical Almanac. This work was started in a newly created
section `Nautical Almanac Unit' attached to the Regional
Meteorological Centre, Kolkata, with Nirmal Chandra Lahiri
(1906-1980) as its first Officer-in-Charge. This newly created
`Unit' published its first issue of `The Indian Ephemeries and
Nautical Almanac for 1958' much earlier in March 1957
accompanying the issues of `Rashtriya Panchang' in English and 11
other Indian languages, so that the Panchang makers all over India
may use for annual publications for the succeeding year. These
annually published Ephemeris and Panchangs contain the actual
timings of all the forthcoming celestial events which are useful
for the scientific, social and religious purposes. Since 1979, the
said `Unit' became an independent institution, `Positional
astronomy Centre' with a Director as its head. This `Centre' in
its turn continues to publish the Panchang annually, but at an
increased number of Indian languages to 14, apart from the
Ephemeris. That the Indian almanac makers can now confidently
publish correctly predicted timings of the celestial events
through their Panchang or Panjika, is the consequence of untiring
effort of the calendar reformers like Chandra.\\

\section{Achievements}
The transformation of an inquisitive boy Chandra, of a remote
village of united Bengal, into a world class astronomer, is itself
a great achievement. He had to discharge his domestic
responsibilities as the father of four children and professional
duties as a staff of Government Treasury, while he voluntarily
engaged himself as an untiring observer of the night sky with all
serious commitments of a professional astronomer. He made
observations, initially, on the celestial incidents like the
shooting stars or meteors, apparition of comets, occurrence of
eclipses and published his observational reports on these in the
national journals. But, lateron his collected data from his
observations on variable stars were of so precious scientific
value that these were reported in the international astronomical
journals. His collected observational data were used by the 
professional astrophysicists for the study and research on 
various physical characteristics of stars. Thus the
night-sky watcher Chandra stood out more gloriously as an 
astronomer than his any other role of the family head or Treasury
Officer.

It was not easy for the general people to realize and appreciate
the quality of academic work, on such rarely studied subject
astronomy, accomplished by Chandra. So, only when the news
concerning a 6.25-inch telescope would be lent to him, was
conveyed by the AAVSO through the letter dated 12 August 1924 by
Leon Campbell, then the enthusiastic goldsmith community of Bengal
accorded overwhelming felicitation, in 1925, to its fellow member
Chandra.

However, Chandra was recognized, time and again, as a world-class
important observer from the East by the very professionals for
regularly contributing observational data on the variable stars.
On request, he used to contribute these data to the astronomical
observatories of Europe and America for the publication in their
respective bulletins and journals and use of the professionals as
well. His observational findings were so unique and fundamental
that he received many spontaneous appreciations and acclamations
from the world class astronomer. One such appreciation, cited
below, was accorded by Harlow Shapley (1885-1972) of Harvard
College Observatory, through a letter dated 12 December 1950:

``The American Association of Variable Star Observers, with
Headquarters at the Harvard Observatory, is honoured to salute you
as one of its important contributors from abroad ..."

He was even rewarded by getting elected as the honourary member of
several learned societies and enlisted as the subscriber of their
respective bulletins or journals published by these institutions.
As of 1953, Chandra was a member of each of several such
societies, a few of which are listed below:\\
      1. American Association of Variable Star Observers, Harvard College
          Observatory, Cambridge, 38 Mass., U. S. A.\\
      2. Association Francaise de Observateurs d$^{\prime}$Eloite Variablese, a`l'
           Observatoire d$^{\prime}$Lyon, France.\\
      3. British Astronomical association, London, England.\\
      4. American Museum of Natural History, New York, U. S. A.

Perhaps `Officer d$^{\prime}$Academie, Republique Francaise' was the only
reward, as such he received from the Government of France through
its Consulate General of Kolkata, on 1 August, 1928. The Consulate
General sent to Chandra a diploma and a badge related to his
reward along with the letter:\\

Consulate General de la Republique Francaise, Calcutta

{\it Dear Sir,\\

In continuation to my letter dated 26-3-28, I have the honour to
inform you that the Ministry of Education has decided to confer
upon you the distinction `Officer d$^{\prime}$Academie'.

You will find herein enclosed the Bravet and the badge of this
distinction for which I shall be obliged to receive a receipt.

I am pleased to convey to you my best congratulations for the
token that has been granted to you in recognition of your valuable
services to the Observatory of Lyon.

~~~~~~~~~~~~~~~~~~~~~~~~~~~~~~~~~~~~~~~~~~~~~~~~~~~~~~~~~~~~~~~~~~~~~~~~~~~~~~~~~~~~~~~~~~~Yours
faithfully,

\noindent Mr. R. G. Chandra
~~~~~~~~~~~~~~~~~~~~~~~~~~~~~~~~~~~~~~~~~~~~~~~~~~~~~~~~~~~~~~~~~~~~~~R.
Lazonies,

\noindent Bagchar,
Jessore~~~~~~~~~~~~~~~~~~~~~~~~~~~~~~~~~~~~~~~~~~~~~~~~~~~~~~~~~~~~~~~~~~Consulate
General for France}\\

Having received the said award, Chandra became so much complacent
that he used to write O. A. R. F. (`Officer d$^{\prime}$Academie, Republique
Francaise') after his name. Also, inspired by the title, `A
Village School master', of a poetry written by Oliver Goldsmith (1730-1774) 
in his school text book, he preferred to be introduced as `a Village Astronomer'.

Chandra wrote quite a few books on astronomy in Bengali. Out of
these only `Dhumketu' (The Comet) was published in 1953 during his
lifetime by the Puthipatra (Calcutta), Pvt. Ltd. The second one
`Tara Chiniber Upaya' (How to recognize the stars) was published
in 1996 by the Bangiyo Bijnan Parishad, Calcutta and the rest,
`Sourajagat' (The Solar System), `Nakshatra Jagat' (The World of
Stars) and `Sabita o Dharani' (The Sun and The Earth) remained
unpublished.\\

\section{Relevance of Chandra}
Even after leaving his ancestral house at East Pakistan after 1947
and taking refuge in parted India, Chandra was able to continue
his activities as an observer astronomer until 1954. In course of
time, due to old age, he gradually became incapacitated for
performing observation through his telescope. Thus his
interactions with the astronomers all over the world, to whom he
was so well known, began to wane. Consequently his image as an
active astronomer squeezed into the physical presence within his
locality at Durgapally. The authorities, teachers, students and
guardians of his neighbouring school Barasat Satyabharati
Vidyapith were the only people remained for interaction by paying
respects on his every birth-anniversary.

During his lifetime, Chandra as an astronomer, was known
to a very few people of his country. Even most of those who knew
him, could not realize his worth as an astronomer. In this
respect, Prof. Apurba Kumar Chakrabarty, a mathematics teacher of
Mahisadal Raj College, Midnapore, was perhaps the first person
known, to assess his worth. Through an article published in
`Modern Review'$^{13}$, he narrated how a person without formal
education, financial assistance and any kind of patronage can
achieve excellent knowledge by observation for the study of
astronomy. Such knowledge was essential for research work of the
contemporary world-class astronomers.

Another scholar, Prof. Amulya Bhusan Gupta, Department of Physics,
Indian Statistical Institute, Calcutta could fully realize the
importance of work done by his next door neighbour Chandra. Prof.
Gupta had ample opportunity to be acquainted with the life and
work of his distinguished neighbour for years together. So, after
the demise of Chandra in 1975, a competent person like Prof. Gupta
recorded an obituary which was published in `Sakswar'$^{14}$, the
magazine of Barasat Satyabharati Vidyapith. From the said
obituary, it reveals that Prof. Gupta made some futile attempts to
draw the attentions of concerned authorities for according
national recognition to such an extraordinary work done by an
ordinary `Village Astronomer'.

The extraordinary achievements of Chandra were
brought into limelight more widely than ever before during the
last two decades of the 20th century by Dr. Ranatosh Chakrabarty,
Surendanath College, Calcutta. Dr. Chakrabarty, also another
neighbour of the former, contributed many articles on Chandra in
the dailies and journals published from India and abroad. He also
presented talks on the same subject through the radio and TV
transmissions, besides delivering invited lectures at different
educational institutions at Calcutta and its neighbourhood. Dr.
Chakrabarty and Prof. I. B. Sinha, a retired professor of
Physics, West Bengal Government Colleges, also a neighbour of the
former, jointly edited in 1985 the second edition of the book
`Dhoomketu' (The Comet) which was originally authored by
Chandra. Dr. Chakrabarty also edited another book
written by Chandra entitled `Tara Chiniber Sahaj Upay'
(The easy means to recognize stars) and was published by the
`Bangiyo Bijnan Parishad', Calcutta, in 1996. Also, Dr.
Chakrabarty wrote himself a book on the life and works of the
village astronomer Chandra published by `Puthipatra'.

In order to perpetuate the memory of the astronomer, Dr.
Chakrabarty and Prof. Sinha instituted the `Radhagobinda Memorial
Society' at Nabapally, Barasat which is the home town of these
three. The society organized several lectures and exhibitions at
schools and colleges in and around Calcutta. During this period, a
good number of articles on the achievements of Chandra were
contributed to the popular journals of Bengal by renowned writers
like Arup Ratan Bhattacharyya$^{15}$, Amalendu Bandyopadhyay and others.

As an impact of these efforts, very significant steps were taken
by different learned institutions of Calcutta. Birla Industrial
and Technological Museum, in collaboration with Indian
Astronomical Society and Radhagobinda Memoral Society,
organized an exhibition on the astronomer for a period of about
two weeks from 18 July 1989. This exhibition was accompanied by a
few lectures and display of the historic 3-inch telescope of
Chandra. In this exhibition, numerous panels depicting available
pictures of Chandra, photo copies of correspondences made
by renowned world-wide astronomers with him and many other papers related
to his astronomical activities were also exhibited.

The West Bengal State Book Board published (March 1990) a
special issue of `Ganit Charcha', a quarterly journal on
mathematics, to commemorate the contributions of Chandra made for
the enrichment of astronomy. A section of eminent scholars of the
subject highlighted in this special issue about his humble life
and role in collecting astronomical data as a lone contemporary
observer from Eastern India. The Editor rued in his editorial that
the birth centenary year 1978 of Chandra passed unnoticed
from his own countrymen. Usually such an occasion offers the
populace an opportunity to be acquainted well about the life and
contribution of a celebrity to the society. From this point of
view, the Editor expected that the said special issue may
compensate, to some extent, the advantages of the missed centenary celebration.

The enchanting scientific works of Chandra also attracted
attention of many people from Bangladesh. At the onset of 21 century, 
the authors such as Fakray Alam (Dainik Bhorer Kagaj,
02.01.2000), Naimul Islam Apu (Dainik Pratham Alo, 18.07.2004) and
Rafikul Islam Sujan (Dainik Samakal, 16.07.2006) contributed
several articles on R. G. Chandra in the news papers. Amongst these, 
Apu has also authored a book entitled `Banglar Jyotirbijnani Radhagobinda 
Chandra'$^{16}$. In order to write the said book, Apu
visited both the ancestral house of Chandra at Bagchar, Bangladesh
and his last residence at Barasat, India. In quest of information
about the `village astronomer', Apu met his grandson Sisir Kumar
Chandra and one of the authors of the present article Sudhindra
Nath Biswas, a retired and National Awarded teacher of Barasat
Satyabharati Vidyapith. During his visit, Apu recorded a short
speech of Biswas on Chandra and arranged for transmission through
Bangladesh Radio (perhaps in 2006).

Another enthusiastic scholar Prof. M. A. Aziz Mia, a retired college
teacher from Bangladesh, had a similar mission. He wrote a scholarly article about Chandra  
in the special issue of `Telescope-2' published from 7, Dhanmundi,
Dhaka - 1205 on the occasion of International Year of Astronomy,
2009.\\

\section{Epilogue}
To end the discussion, let us recall the wellknown proverb ``some
are born great, some achieve greatness and greatness are thrust
upon some others''. The village astronomer Chandra certainly
belonged to the second category with respect to his greatness.
Indeed he overcame his limitations, both as an academician and an
observational astronomer by sheer zeal, enthusiasm and
perseverance. As a variable star observer, his meticulously
collected data helped the professional astronomers to develope
theories regarding the variability of the stars while as a comet
observer he not only observed the beauty of the comets, but also
showed his scientific attitude by noting down the day to day
specific location of the comets. Being undettered by missing the
rare honour of a nova discovery, he went on performing the tedious
job of night sky observation. This proves that in respect of
motivation and dutifulness he was at per with the professional
astronomers. This marks the upliftment of Chandra which he
achieved through his own effort and nothing else. In this context 
it will be very relevant to quote ``Radha Gobinda Chandra's (1878 - 1975) 
introduction to modern astronomy came not through his European employers, 
but through Bengali texts, and he became a researcher, though a modest one.''$^7$ 
This will serve as a beacon of light to future generations of astronomers - both
amateur as well as professional.\\

\section*{APPENDIX I}
Kalinath Mukherjee was born in a middle-class Brahmin family of
the village Jaidia in the district of Jessore. He had his
undergraduate education from Krishnanagar College through which he
was conferred the degree of B. A. by the University of Calcutta in
1872 with honours in mathematics, philosophy and Sanskrit. He
continued his studies with the subject Law and was again conferred
the degree of B. L. by the same university in 1873. While he was a
college student, Mukherjee had the opportunity to come in close
contact with Sir M. J. Herschel, M. A., and Bar-at-Law. Herschel
was the then District and Session Judge of Nadia and was posted at
its Headquarter Krishnanagar. He was the descendent of a world
famous family having great astronomical background as he was none
other than the son of Sir John Herschel (1792 - 1871) and the
grandson of Sir William Herschel (1738 - 1822) who were the two
pioneer astronomers of their times in their own rights. Mukherjee
had the privilege of getting immense inspiration and guidance from
such a person as M. J. Herschel for the study of astronomy.

After obtaining the B. L. degree, Mukherjee started to practice
Law at the bar in his home district town Jessore. Soon he
succeeded in his profession and became an eminent lawyer. Yet,
astronomy remained the subject of his great interest and devotion.
He used to visit his native village Jaidia, far away from the
district town, in every week-end for sky observation. There he
constructed a small observatory and named it as the Ripon Palli
after the name of Lord Ripon who was the Viceroy of India (1880 - 1884). 
Mukherjee used to contribute articles on
astronomical topics in the leading journals of Bengal and became a
popular writer of the same topics. Subsequently he published in
1901 a sky map entitled `Bhagola Charitam' written in Sanskrit. He
authored another book under the title `Tara' (The Star) in Bengali
and published it in the year 1902. Also in 1905 he authored a book
entitled `Popular Hindu Astronomy'$^{17}$ written in English with
extensive quotations from classical Sanskrit literature, viz.
Vedas, Ramayana, Mahabharata, Puranas etc. Mukherjee was also so
much committed as an amateur astronomer that at the end of day's
professional duties, he used to arrange meetings at his residence.
The said meetings were attended by the persons interested in
astronomy to discuss the topics related to the celestial bodies.\\

\section*{APPENDIX II}
The discovery of Nova Aquilae was first reported in Nature in
its 13 June 1918 (page 285) issue by F. W. Dyson. Actually, Dyson
reported on the basis of intimations he received on the
independent discoveries of the Nova made by several observers. In
his report Dyson states : {\it ``Apparently, the first observation
was made by Miss Grace Cook at Stowmarket when on the watch for
meteors at $9^h~30^m$~P.M., G.M.T. on June 8"}. After four months,
John Evershed claimed$^{18}$: {\it ``In Nature of June 13, I note
that the earliest observation of Nova Aquila in Enland was made by
Miss Grace Cook at $9^h~30^m$~P.M., G.M.T. on June 8 and the
magnitude was estimated as equal to Altair. In India the star was
seen and recognized as a nova about five hours earlier by Mr. G.
N. Bower in Madras, who has sent me his original notes made at 10
P.M., Indian Standard Time on June 8 (corresponding with 4-30
P.M., G.M.T.)"}.

Evershed must had no knowledge about the observation of Nova as a
`new star' made by Chandra from a further eastern longitude at
Bagchar even one night ahead of any other observer on the night of
6-7 June, 1918, at an earlier stage when its gradually increasing
brightness was only 6 magnitude. On the night of 8-9 June 1918,
Chandra made his second successive night observation while Mr.
Bower of Madras ($13^0~4^{\prime}$~N, $80^0~17^{\prime}$~E) and
Miss Cook of Stowmarket, England both made their respective
`first' observation on the `new star' at more brighter stage than
the previous night. Yet, this remarkable achievement remained
unrecognized due to his ignorance of the formalities involving
astronomical discoveries.\\

\section*{APPENDIX III}
In a letter dated 29 May, 1920, the Secretary of AAVSO Mr. How and
O$^{\prime}$Eaton appreciated Chandra by writing - {\it ``I want to thank
you very much for the splendid list of observations of variable
stars ..."}$^1$

Harlow Shapely, Director of Harvard College Observatory, sent a
letter (June 20, 1922) to Chandra with the words - {\it ``May I
add a personal word of congratulation for the good work you have
been doing in the observation of long period variable stars. Your
longitude is of considerable importance in this work."}$^1$

Felix de Roy, Director, Variable Star Section, British
Astronomical Association, expressed his desire to meet Mr.
Chandra, in a letter dated 27 December 1923, by writing - {\it ``I
seize this opportunity in saying that your excellent observations
and remarks are always much valued by this section. ... I should
be personally pleased to meet you if ever you were to cross to
Europe."}$^1$

In a letter dated 29 March 1935, the Recording Secretary of AAVSO
Leon Campbell congratulated Mr. Chandra. The letter goes as - {\it
``I am pleased to receive such a splendid report of observations
as made during January  and congratulate you on the excellence of
observations"}. In the same letter, Mr. Campbell wrote - {\it ``I
am pleased to acknowledge your postal card of the 16 January
concerning the first estimates of Nova Herenlis. I am glad to note
that you have kept watch on this star and have secured such a
continuous series of observations. The fluctuations noted inn the
star certainly bear real and these have not ceased even up to the
present time"}. So, this letter of Campbell reveals that Chandra
estimated the variability of Nova Herenlis for the first time$^1$.

In his article `The Role of the Amateurs in Variable Star
Astronomy', Leon Campbell has remarked - {\it ``In foreign
countries we have Radha G. Chandra, official of Bagchar, India.
Mr. Chandra, now in his sixtieth year, who has been aiding in the
variable star work since 1919, has accumulated probably more
observations on variable stars than any other AAVSO foreign
observer, well over 50,000."}$^1$\\

\section*{APPENDIX IV}

{\it My dear Mr. Chandra,

For a longtime I have been very desirous of securing for you the
use of a larger telescope than you have. At last this seems about
to be realized. Our patron and friend Mr. C. W. Elmer of N. Y. has
turned over to the Association his $6.25$ inch lens in tube with
finder and oculars and cradle claps for attaching the tube to a
mounting. The lens is a very good one and should enable one to see
much fainter objects than with a three inch instrument.

Now if you can see your way clear to provide some sort of
mounting, either temporary or permanent, the telescope committee
is willing to let you have the loan of this splendid equipment as
described above.

... Just how to best arrange for such a loan is the serious
question. As long as you live and will keep the instrument in
reasonable use for AAVSO. For variable star observing, the
telescope can be considered as virtually yours. The difficulty
comes in the case of your death. What assure can be had that the
equipment would be resorted to the Association, either to some
other observer in Asia, Europe or U. S. A.? This is what bothers
us. The equipment is valued at 500 dollars and as long as you keep
it in good use, we shall feel well repaid for our efforts in
lending to you.

I might suggest that your own three inch would make a very
desirable additional finder, especially if you have no circles at
first. The Association could defray the initial cost of
transportation asking you to repay the Association as you could.

If you decide that the loan of this equipment is practical, that
you can provide some sort of mounting for the present at least and
will agree to use it exclusively for AAVSO observing, will take
good care of it and reimburse the Association later for the
transportation expenses and provide for its return to the
Association or its authorized agent upon your demise or inability
to make further use of it. Let me know and I shall start steps for
having it sent to you at once.

With best wishes and kind regards, I am

Faithfully yours,

Leon Campbell

Chairman of Telescope Committee.}\\

\begin{center}
ACKNOWLEDGEMENT\\
\end{center}
We all thank Mr. Sisir Kumar Chandra, grandson of R. G. Chandra for helpful discussions. 
Thanks are also due to Prasenjit Basu, The Future Institute of Engineering
and Management, Kolkata and Farook Rahaman, Jadavpur University,
Kolkata for their technical help in different forms that improved
the quality of the article a lot. SR is personally grateful to 
Prof. J. V. Narlikar for several suggestions and also to the
authority of IUCAA, Pune for providing Visiting Associateship
under which a part of this work was carried out.\\

\begin{center}
 REFERENCES
\end{center}

\begin{enumerate}

\item{} Ranatosh Chakrabarty, `Jyotirbijnani Radhagobinda'  
(Puthipatra, Kolkata, 1999).

\item{} Radhagobinda Chandra, `Dhumketu' (Puthipatra, Kolkata, 1985).

\item{} Catalogue of Cometary Orbits (eds. B. C. Marsden and C. V. Williams), Smithsonian Astrophysical Observatory, Cambridge, U.S.A. (1999).

\item{} Nature, Vol. 107, No. 2700, p. 694 (1921).     

\item{} Monthly Reports and Annual Reports of the American
Association of Variable Star Observers, p. 133 (1926).

\item{} Memoirs of the British Astronomical Association, vol. XXVIII, Table C. (1929).

\item{} Rajesh Kochhar and Jayant Narlikar, `Astronomy in India: 
Past, Present and Future' (IUCAA, Pune and IIA, Bangalore, 1993).

\item{} Astronomy of the 20th Century : Otto Struve and Velta Zeberss (Macmillan Co., New York, p. 354, 1962). 

\item{} Journal of the British Astronomical Association 34, 241.

\item{} Ibid, 42, 178.   

\item{} Ibid, 44, 157.   

\item{} Ibid, 45, 407.  

\item{} Modern Review (June, 1971, pp 445-449).

\item{} Amulya Bhusan Gupta, `Nakshatrabid Radhagobinda', 'swaksar' (Barasat Satyabharati vidyapith Magazine), p. 2-7 (1975). 
      
\item{} Naimul Islam Apu, `Jyotirbid Radhagobinda Chandra', Tamralipi, Dhaka - 1100, Bangladesh (2008). 

\item{} Arup Ratan Bhattacharya, `Radhagobinder Aprokasito Pandulipi', Anandamela, 12 July (1989). 

\item{} Kalinath Mukherjee, `Popular Hindu Astronomy', published by Nirmal Mukherjee, 18, Deshapriyo Park, Kolkata - 700 026 (1905). 

\item{} Nature, p. 105, 10 October (1918).

\end{enumerate}

\end{document}